\documentclass[aps,pra,reprint,showpacs,floatfix,amsmath,amssymb,10pt]{revtex4-1} 

\usepackage[pdftex]{graphicx}       

\usepackage{epstopdf}
\usepackage{txfonts}        

\usepackage[plainpages=false,pdfpagelabels,colorlinks]{hyperref} 
\hypersetup{colorlinks,citecolor=black,filecolor=black,linkcolor=black,urlcolor=black}
\hypersetup{colorlinks,citecolor=blue,filecolor=blue,linkcolor=blue,urlcolor=blue}

\usepackage{amsmath}				

\newcommand{\scA}{\mathcal{A}}

\newcommand{\scP}{\mathcal{P}}

\newcommand{\aho}{a_\text{ho}}
\newcommand{\rr}{{\mathbf r}}	
\newcommand{\RR}{{\mathbf R}}
\newcommand{\Amat}{{\mathbf A}}
\newcommand{\xx}{{\mathbf x}}
\newcommand{\kch}{^{(k)}}
\newcommand{\phik}{\phi_k \left(\xx\kch \right)}
\newcommand{\dd}{\mathrm{d}}

\newcommand{\hb}{\hbar}
\newcommand{\oth}{^\text{th}}

\newcommand{\eref}[1]{Eq.~\eqref{eq:#1}}
\newcommand{\fref}[1]{Fig.~\ref{fig:#1}}
\newcommand{\sref}[1]{Section \ref{sec:#1}}

\begin{document}


\title{Coupled pair approach for strongly-interacting trapped fermionic atoms}

\author{C. J. Bradly} 
\author{B. C. Mulkerin} 
\author{A. M. Martin}
\author{H. M. Quiney}
\affiliation{School of Physics, University of Melbourne, Victoria 3010, Australia}

\date{\today}
\begin{abstract}
We present a coupled pair approach for studying few-body physics in harmonically trapped ultracold gases. The method is applied to a two-component Fermi system of $N$ particles. A stochastically variational gaussian expansion method is applied, focusing on optimization of the two-body correlations present in the strongly interacting, or unitary, limit. The groundstate energy of the four-, six- and eight-body problem with equal spin populations is calculated with high accuracy and minimal computational effort. We also calculate the structural properties of these systems and discuss their implication for the many-body ultracold gas and other few-body calculations.
\end{abstract}

\maketitle

\section{Introduction}

Ultracold atomic two-component Fermi gases under harmonic confinement have become an important field of study for fundamental quantum phenomena. The tunability of the interspecies $s$-wave scattering length -- the dominant interaction channel -- makes these systems ideal for exploring the strongly interacting, or unitary, regime at the BEC-BCS crossover where the scattering length diverges and becomes the dominant length-scale in the system \cite{Shin2008,Chin2010,Braaten2006,Zwierlein2006,Blume2012}. Many-body calculations based on perturbative methods fail at unitarity due to the divergence of the scattering length \cite{Bloch2008,Giorgini2008}. Alternative techniques involving Monte Carlo integration and effective interactions \cite{Gilbreth2012, Blume2007,Mukherjee2013} require accurate knowledge of the nodal surfaces to be used as references for antisymmetric wavefunctions for fermionic systems. Density functional theory requires an accurate energy functional for the study of many-body systems \cite{Bulgac2002,Xianlong2006,Papenbrock2005}. Studies of few-body systems provide benchmarks for optimization and refinement of these many-body calculations and future experiments \cite{Werner2006,Werner2006a,Kestner2007,Daily2010,Liu2009,Liu2010,Stoeferle2006}. 

Few-body calculations can also be directly applied as atom traps become more sophisticated and offer the possibility of trapping only a few atoms in one trap \cite{Serwane2011} or a few atoms on each site of an optical lattice \cite{Bakr2009,Stoeferle2006}. Two-body correlations have been observed to play an important role in these systems \cite{Zuern2013}. Extensions to more complex systems involving the application of external fields expands the known set of universal relations \cite{Mulkerin2012,Mulkerin2012b}. 

The in-principle exact calculation of harmonically trapped few-body systems with zero-range $s$-wave interactions has been greatly extended in recent years. The exact wavefunction and energy spectrum for two unlike atoms in a trap was found by Busch {\em et al.}~\cite{Busch1998}. Knowledge of the two-body system has spurred calculations of the three-body problem using the adiabatic hyperspherical method \cite{Werner2006,Kestner2007,Liu2010}. Exact diagonalization using the stochastic variation of a correlated gaussian basis has enabled calculation of energetics and structural properties of up to six trapped fermions for finite range interactions \cite{Daily2010,Blume2011}. However, as the number of particles increases the Hilbert space grows exponentially and exact diagonalization becomes intractable. The challenge is to extend these calculations as far as possible beyond the two- and three- body problem towards the many-body regime.


In this work we present an improved methodology based on the stochastic variation of a gaussian basis \cite{Mitroy2013,Suzuki1998,Hiyama2003} that allows calculation of energy levels and structural properties of a two-component Fermi system with up to \smash{$N=8$} atoms. The atomic hyperfine states which form the two components are treated as two arbitrary spin-$1/2$ states and the associated statistics only allows interactions between unlike particles. In this work we restrict ourselves to the case of even $N$ with equal population in each spin state, i.e.~\smash{ $N_\uparrow = N_\downarrow = N/2$}, as proof of concept of the approach. The groundstate of this system has zero total orbital angular momentum and spin, simplifying the calculation. The extension to more general cases is straightforward but requires greater computational effort.

The key idea of the coupled pair approach is to consider only the essential correlations. Interactions only occur between two unlike fermions and all other correlations are captured in the non-interacting correlations between dimers whose behavior is governed by the trap. This problem can be solved using variational methods but the optimization procedure is only applied at the two-body level simplifying the calculation for larger $N$ where exact diagonalization is normally intractable. 
With these considerations we use a gaussian expansion of the relative wavefunction and stochastic optimization to calculate the groundstate energy and structural properties of up to \smash{$N=8$} fermionic atoms in a harmonic trap. 

\sref{GenApproach} outlines the basic formalism for $N$ trapped fermions. In \sref{FourBody} we discuss the \smash{$N=4$} problem in detail and introduce the coordinate channels and their importance for the few-body problem. \sref{SixAndEight} extends the method to \smash{$N=6$} and \smash{$N=8$} and outlines the details of the coupled pair approach. Then, in \sref{Structural} we calculate the single-particle density and pair correlation functions and discuss their importance to the many-body system. Finally, in \sref{Conclusion} we summarize and discuss the extension of the approach to more general cases.

\section{\texorpdfstring{$N$}{N}-body problem and general approach}
\label{sec:GenApproach}

The Hamiltonian  for $N$ harmonically trapped atoms of equal mass $m$ is
\begin{alignat}{1}
H=\sum_{i=1}^N  \left ( -\frac{\hbar^2}{2 m} \nabla^2_i +\frac{1}{2}m\omega^2 r_i^2 \right ) + \sum_{i<j} V(\rr_i-\rr_j),
\label{eq:Hamil_main}
\end{alignat}
where $\rr_i$ is the position of particle $i$, $\omega$ is the trapping frequency and $V(\rr)$ is the interparticle interaction potential. The sum is restricted to interactions between unlike fermions and interactions between more than two particles are omitted. Here we only consider the equal mass case but the following discussion can be generalized to arbitrary masses. Firstly, the center-of-mass motion is decoupled from \eref{Hamil_main} and solved separately, whereupon we may assume that it is in the groundstate with energy \smash{$E_\text{cm}=1.5 \hbar\omega$} and wavefunction
\begin{alignat}{1}
\psi_{\text{cm}}(\RR)=\frac{N^{3/4}}{\pi^{3/4}\aho^{3/2}}\exp\left(-\frac{R^2}{2\aho^2/N}\right),
\end{alignat} 
where \smash{$\RR=\sum_i \rr_i/N$} is the center-of-mass coordinate and \smash{$\aho=\sqrt{\hbar/m\omega}$} is the harmonic oscillator length.

All the interparticle interactions are then contained in the relative Hamiltonian and the relative wavefunction is an antisymmetrized function of \smash{$N-1$} relative coordinate vectors. The choice of the relative coordinates is not unique and can be chosen to take advantage of the symmetry of the system and reduce the complexity of the problem. Different sets of coordinates can represent different channels in which the particles are correlated, and we can include multiple channels in our ansatz for the relative wavefunction 
\begin{alignat}{1}
\psi_\text{rel}\left(\rr_1,\ldots,\rr_N\right) = \sum_k \scA\, \phik,
\label{eq:RelativeWavefxn}
\end{alignat}
where $\phik$ is an unsymmetrized wavefunction of the $k\oth$ channel and the set of coordinates for each channel is expressed as a supervector $\xx\kch$. The operator $\scA$ projects these channel wavefunctions onto the correct antisymmetric space.

The problem of two trapped fermions has been solved analytically for the zero-ranged Fermi pseudo-potential and reproduces the \smash{$1/r-1/a_s$} cusp in the relative wavefunction \cite{Busch1998}, where $a_s$ is the $s$-wave scattering length. Unfortunately, these analytic solutions are unsuitable for use in the $N$-body problem, due to the difficulty of treating the many cusps that occur and because the non-interacting harmonic oscillator states have poor convergence at unitarity. Instead we use a gaussian basis to expand the \smash{$\phik$} of each channel
\begin{alignat}{1}
\phik &= \sum_{i_1} \dots \sum_{i_{N-1}}c_{i_1\ldots i_{N-1}} \prod_{j=1}^{N-1} \exp \left( -\frac{{x_j\kch} ^2}{2 \alpha_{j\,i_j}^2} \right),
\label{eq:GaussianBasis}
\end{alignat}
where the $c_{i_1\ldots i_{N-1}}$ are expansion coefficients, and the $\alpha_{j\,i_j}$ are the widths of the gaussians, which serve as variational parameters for the model. Each of the relative vectors $\xx\kch_j$ represents a correlation within the problem and are treated independently. Hence, each basis term is separable in the \smash{$N-1$} vectors, as represented by the product in \eref{GaussianBasis}. In general, this product can have more terms corresponding to any number of correlations, but as explained below, we choose the same number of correlations as relative coordinates. The basis size is determined by the number of terms in the sums and can be expanded to improve the solution in accordance with the variational principle \cite{Mitroy2013}. The groundstate of the equal spin component problem has zero relative angular momentum so our ansatz needs no explicit angular component. The gaussian basis functions are not orthogonal but they are simple to manipulate and are effective at replicating correlations at any length scale, including the short-ranged interparticle interactions, while also being efficient enough to scale to larger systems. 

In the unitary limit the $s$-wave scattering length $a_s$ diverges and becomes the only important length-scale associated with the interparticle interaction.
The details of the short-ranged interparticle interaction potential $V(\rr)$ are unimportant, provided it can support a single bound state. We chose a gaussian basis for its flexibility and ability to access all length scales so we also choose a gaussian potential
\begin{alignat}{1}
V(\rr)=V_0\exp\left(-\frac{r^2}{2r_0^2}\right),
\label{eq:GaussianPotential}
\end{alignat}
where, for any width $r_0$ there is a depth $V_0$ such that the potential supports a single resonant bound state and has a divergent scattering length corresponding to the unitary limit. In the limit \smash{$r_0/\aho \to 0$} the gaussian potential tends towards a regularized contact interaction and the bound state is well behaved. Moreover, it simplifies the calculation of matrix elements when using a gaussian basis and the appropriate values of $r_0$ and $V_0$ can be found with elementary scattering theory. Universal properties only emerge in the true zero-range limit but the values of $r_0$ considered here are sufficiently small for the properties of the system to be considered very close to the true groundstate, for \smash{$r_0/\aho \to 0$}.  There are $(N/2)^2$ possible interacting pairs for the equal spin component system and we include all of them in the Hamiltonian \eqref{eq:Hamil_main}.

The product of gaussians in \eref{GaussianBasis} can be conveniently expressed with an \smash{$(N-1)\times(N-1)$} symmetric matrix \smash{$\Amat^{(k)}$} as
\begin{alignat}{1}
\prod_{j=1}^{N-1} \exp \left( -\frac{{x_j\kch} ^2}{2 \alpha_{j\,i_j}^2} \right) &=  \exp \left( -\frac{1}{2} {\xx\kch}^\text{T} \Amat^{(k)} 
\xx\kch \right),
\label{eq:GaussianMatrixBasis}
\end{alignat}
where the superscript `T' denotes matrix transposition. The matrix elements of all terms in the Hamiltonian \eqref{eq:Hamil_main} with the gaussian potential \eref{GaussianPotential} can be found from these correlation matrices \cite{Suzuki1998}.

By diagonalizing the relative Hamiltonian we not only obtain the energy spectrum but the relative wavefunction. Combining this with the center-of-mass wavefunction we obtain the total wavefunction $\Psi(\xx)$ for the $N$-body problem. 
From $\Psi(\xx)$ we can calculate a general structural property $P(r)$
\begin{alignat}{1}
P(r)=\int\!\!\dd \rr'\,\frac{\delta(r-r')}{4\pi r'^2} \int\!\! \dd^{3N}\xx\, \delta(\rr'-\xx) |\Psi(\xx)|^2,
\label{eq:GeneralDensity}
\end{alignat}
where $\rr$ (and $\rr'$) is a coordinate describing the property of interest and $P(r)$ is normalized to unity. Here $\xx$ is a general set of coordinates such as the center-of-mass plus relative coordinates as defined above or the single-particle coordinates. These quantities are related to the density matrices of the system and are calculated in a similar way \cite{Suzuki1998,Blume2011}. In particular we calculate the single-particle reduced density \smash{$P_1(r)/\aho^{-3}$}, with \smash{$\rr=\rr_1$} in \eref{GeneralDensity} and the (scaled) pair correlation function \smash{$4\pi r^2 P_{12}(r)/\aho^{-1}$}, with \smash{$\rr=\rr_1-\rr_2$} in \eref{GeneralDensity}. $P_1(r)$ is the density of either spin species and $P_{12}(r)$ is the probability of finding a pair of opposite spin fermions of size $r$.

\begin{table}[b!]
\caption{The coordinates for the three linearly independent channels used in the \smash{$N=4$} problem. The reduced mass of all coordinates is $\mu=4^{-1/3}m$.}
\begin{tabular}{l|ccc}
\hline \hline
$k$ & $\xx_1\kch$ & $\xx_2\kch$ & $\xx_3\kch$ \\ 
 \hline  
K$_1$	& $\sqrt{\frac{1}{2\mu}} \left(\rr_1-\rr_2 \right)$	& $\sqrt{\frac{2}{3\mu}} \left(\frac{\rr_1+\rr_2}{2}-\rr_3 \right)$ &
  $\sqrt{\frac{3}{4\mu}} \left(\frac{\rr_1+\rr_2+\rr_3}{3}-\rr_4 \right)$ 	\\
K$_2$	& $\sqrt{\frac{1}{2\mu}} \left(\rr_1-\rr_2 \right)$	& $\sqrt{\frac{2}{3\mu}} \left(\frac{\rr_1+\rr_2}{2}-\rr_4 \right)$	
  & $\sqrt{\frac{3}{4\mu}} \left(\frac{\rr_1+\rr_2+\rr_4}{3}-\rr_3 \right)$ 	\\
H	& $\sqrt{\frac{1}{2\mu}} \left(\rr_1-\rr_2 \right)$	& $\sqrt{\frac{1}{2\mu}} \left(\rr_3-\rr_4 \right)$	& $\sqrt{\frac{1}{\mu}} \left(\frac{\rr_1+\rr_2}{2} - \frac{\rr_3+\rr_4}{2} \right)$ 	\\
\hline \hline
\end{tabular}
\label{tab:FourBodyChannelCoords}
\end{table}


As seen by the choice of basis, our approach identifies interparticle correlations in the $N$-body problem with the relative coordinates. These `channels' serve the dual purpose of being a separable set of coordinates in which to perform the integration of the Schr\"odinger equation while also directly corresponding to the most important correlations in the problem. Other correlations can be included as off-diagonal elements in $\Amat^{(k)}$. A common approach is to include all \smash{$N(N-1)/2$} two-particle correlations \cite{Daily2010,Rittenhouse2011,Mitroy2013}, but having more variational parameters quickly becomes intractable for $N\gtrsim6$. Furthermore, correlations between like fermions that cannot interact are much less significant at unitarity, and all the important correlations are naturally included in the choice of coordinates. 
By only including \smash{$N-1$} parameters from each channel the calculation of individual matrix elements is more efficient, since the $\Amat^{(k)}$ are diagonal, and the basis we use focuses more on the most important two-particle correlations. Therefore, the choice of coordinates is the key to not only encapsulating the important properties of the $N$-body problem but to make it tractable as $N$ increases. We first illustrate these ideas in the case of \smash{$N=4$}.

\section{Four-body problem: \texorpdfstring{$N=4$}{N=4}}
\label{sec:FourBody}
\subsection{Coordinate channels}

For the four-body problem the relative coordinates can be constructed in two ways, K-type and H-type. K-type coordinates are constructed by iteratively defining the relative vector between the center-of-mass of a subgroup of particles and one extra particle; they are the canonical Jacobi coordinates. Physically, in the four-body problem the K-type channels represent the correlations between a pair and two free particles. H-type coordinates begin by defining two pairs and then the relative vector between the centers-of-mass of the pairs. These channels represent the correlations within two interacting dimers, and then the dimer-dimer correlation. 

There are \smash{$4!=24$} ways to construct any coordinate channel depending on the order in which the particles are correlated and so the four-body problem has 48 possible coordinate channels \cite{Hiyama2003}. However, we do not need to include all channels since the symmetry of the problem will make many of them redundant. 
The antisymmetrizing operator for two spin-up and two spin-down particles is \smash{$\scA = (1-\scP_{13}-\scP_{24}+\scP_{13}\scP_{24})$}, where $\scP_{ij}$ permutes the $i\oth$ and $j\oth$ particles. Throughout this work we have adopted the notation that odd and even indices label the two different spin components. Under the action of the permutation operators in $\scA$ and considering the interaction terms included in the Hamiltonian \eref{Hamil_main} the 48 K-type and H-type channels in \eref{RelativeWavefxn} for the four-body problem reduce to only three linearly independent channels, shown in \fref{FourBodyChannels} and given explicitly in Table \ref{tab:FourBodyChannelCoords}. These channels contain all the correlations required for the four-body problem. The reduced mass \smash{$\mu=N^{-1/(N-1)}$} is the same for all relative coordinates and preserves the volume element in the transformation from single-particle coordinates. The antisymmetrizer also has the effect of ensuring that all possible interacting pairs of particles produce a cusp in the wavefunction, without having to explicitly include them all in the basis. 

\begin{figure}[t!]
\includegraphics[width=.9\columnwidth ]{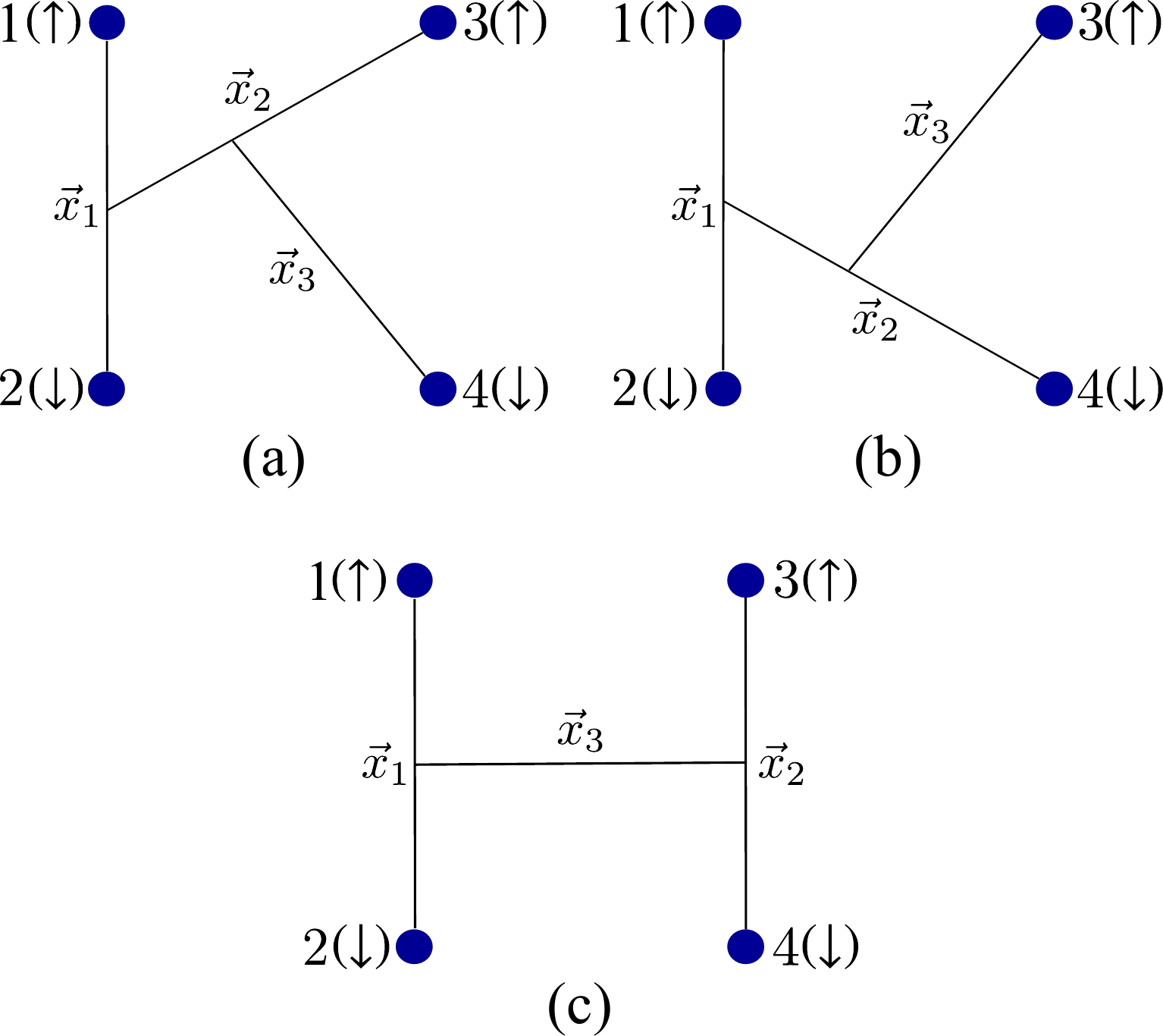}
\caption{The three sets of linearly independent relative coordinates for \smash{$N=4$}. (a) and (b) are conventional Jacobi vectors and (c) is the fully-paired H-type channel. Each line represents the correlation between the centers-of-mass of two subsets of particles. All other channels can be obtained by applying the antisymmetrizer to one of these three channels so they contribute nothing extra to the solution.}
\vspace{-0.5cm}
\label{fig:FourBodyChannels}
\end{figure}

\subsection{Groundstate energy for \texorpdfstring{$N=4$}{N=4}}

For excited states or weaker interactions the correlations of free particles is important and the contributions from the K-type channels must be included. However, for the groundstate at unitarity we intuitively expect that the H-type channel is sufficient.
In each channel the basis is chosen using the stochastic variational method whereby a set of gaussian widths is chosen semi-stochastically and the relative Hamiltonian is constructed and diagonalized. This process is iterated until the lowest eigenenergy converges.
These calculations are repeated for $r_0/\aho=0.05$ to $r_0/\aho=0.01$ and show a linear trend that is extrapolated to the limit \smash{$r_0/\aho \to0$}. Using all three channels from \fref{FourBodyChannels} we achieve a groundstate energy of \smash{$E_\text{G}^{(4)}=3.509(6)\hbar\omega$}, very consistent with previous calculations \cite{Daily2010,Rittenhouse2011}. The uncertainty is in the last digit. With this calculation as a benchmark, we can compare to a calculation using only the H-type channel, for which we obtain \smash{$E_\text{rel}=3.51(3)\hb\omega$}, or a difference of $0.15\%$ from the multiple channel calculation. These calculations are summarized in Table \ref{tab:EnergyResults}.

The H-type channel by itself gives a very good approximation to the true groundstate energy, but as a further test we examine the structural properties of the system at unitarity. In \fref{FourBodyMultiChannelDensity} we plot the scaled pair correlation function \smash{$4\pi r^2 P_{12}(r)/\aho^{-1}$} of the four-body groundstate at unitarity using different combinations of channels. The calculation is performed using only the H-type channel (blue, solid), only the K-type channels (red, dashed), and all three (black, dotted). As shown by the inset, the calculation using only the H-type channel is very close to the calculation using all three channels. The minimal effect of the adding the K-type channels to the H-type channel is to make the pair sizes slightly smaller, reflecting higher-order pair correlations. The K-type channels by themselves are very different from the full calculation, especially at small $r$. This is due to the fact that they do not allow for as many pairs of unlike fermions with a small separation.

These results confirm that the groundstate at unitarity is well-represented by only the H-type channel. That is, the most important correlations in the system are those between two particles interacting via an $s$-wave contact interaction and then between the dimers whereas correlations involving free particles contribute only a small perturbation. Interactions involving more than two particles are also negligible. Therefore, when we consider scaling to larger $N$ in the next section, we seek to discard the insignificant terms and devote computational power to the interacting pairs. 

\section{Extension to higher \texorpdfstring{\smash{$N$}}{N}}
\label{sec:SixAndEight}
\subsection{Coordinate channels and optimization}

In the four-body problem, the different correlations contained in the K-type and H-type channels leads to a clear distinction between the results. For larger $N$ there arise more possible sets or `shapes' of coordinate channels, including hybrids of the generalized K-type and H-type channels introduced earlier and each set allows $N!$ permutations before considering the antisymmetry of the wavefunction. The different coordinate channels can be characterized by the types of correlations they naturally represent and any channel can be viewed as a set of \smash{$N=2$} subsystems. We identify two types of correlations, interacting-pair correlations (IPCs) between two fermions in different spin states and non-interacting correlations (NICs) involving more than two particles. The latter includes correlations between dimers, and correlations between a subcluster (two or more particles) and a single particle. The distinction is important because the IPCs in the relative wavefunction must reproduce the \smash{$1/r-1/a_s$} cusp between any two interacting fermions. Although correlations involving more than two particles can be treated via effective interactions \cite{Petrov2004}, cusps in the NICs are either suppressed by Fermi statistics or are much weaker than the two-body interaction and the correlations are on a larger length scale. These higher-order effects have not been considered explicitly in this work but their effects can be considered to be incorporated in the NICs which are governed primarily by correlations of order $\aho$. For higher $N$ H-type refers only to those channels which have the maximum $N/2$ IPCs. That is, they are constructed by first pairing up all particles then building the correlations between pairs. All other channels are referred to as (generalized) K-type, even if they contain several IPCs.

\begin{figure}
\includegraphics[width=0.9\columnwidth ]{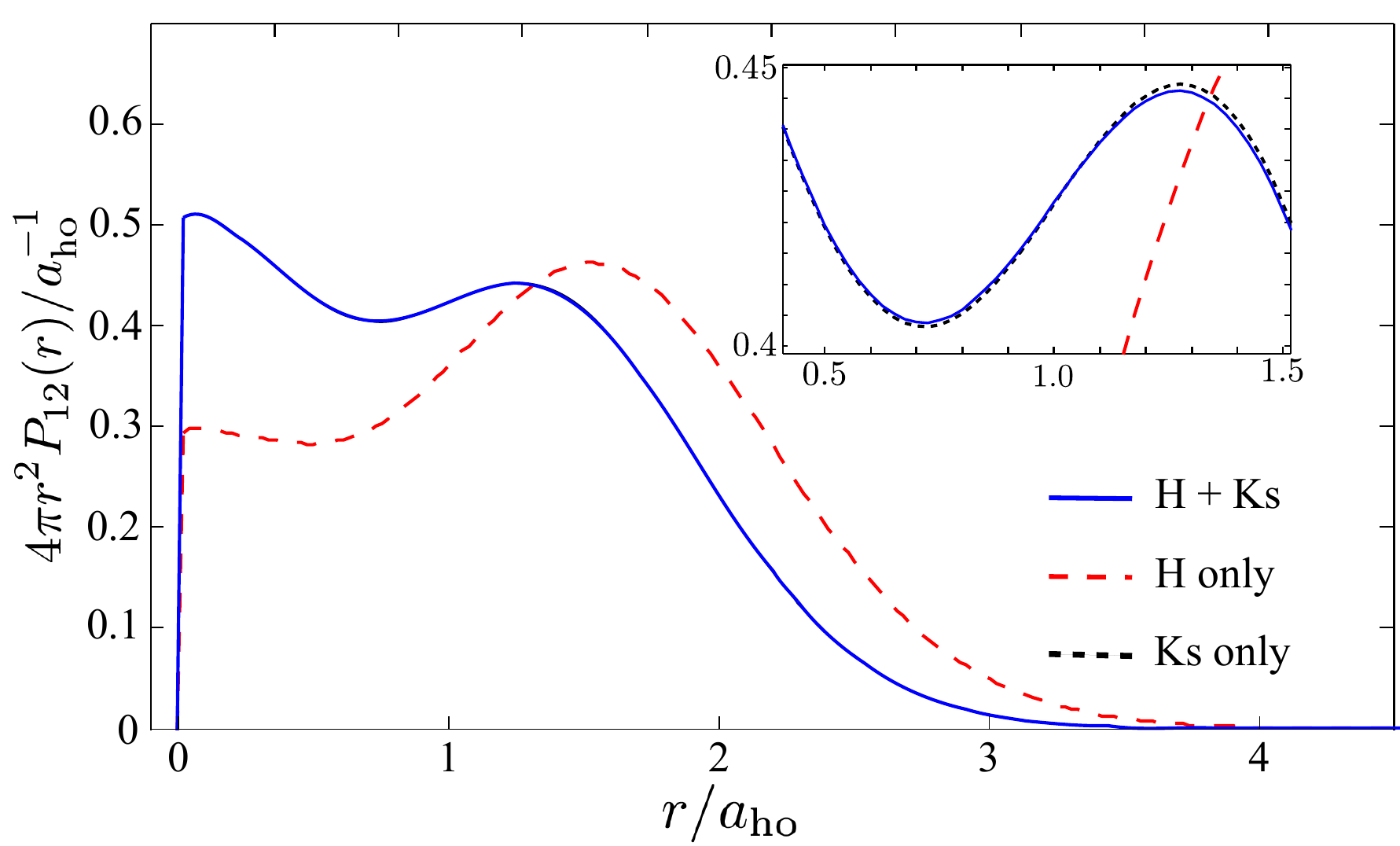}
\caption{The scaled pair correlation function \smash{$4\pi r^2 P_{12}(r)/\aho^{-1}$} of the four-body groundstate at unitarity using different combinations of channels. The calculation is performed for \smash{$r_0/\aho=0.01$} using only the H-type channel (blue, solid), only the K-type channels (red, dashed), and all three (black, dotted). The inset demonstrates the robustness of the calculation using only the H-type channel. For each calculation the basis used was the same size for each channel so the H-type only calculation had the smallest basis.}
\vspace{-0.5cm}
\label{fig:FourBodyMultiChannelDensity}
\end{figure}

\begin{figure}[t!]
\includegraphics[width=\columnwidth]{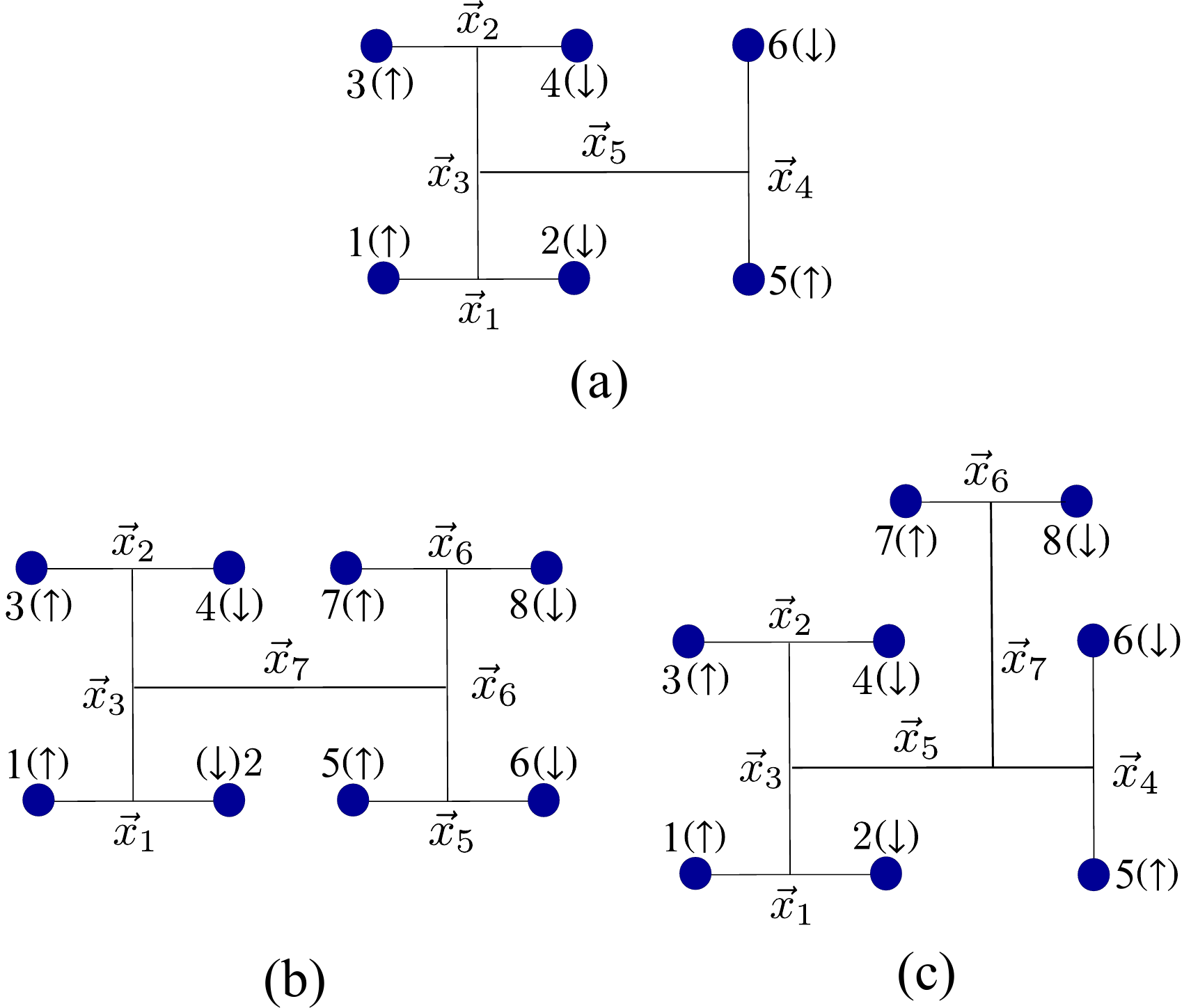}
\caption{(a) The H-type channel for the \smash{$N=6$} problem, (b)-(c) the H-type channels for the \smash{$N=8$} problem. All other fully-paired channels can be obtained by applying the antisymmetrizer to one of these channels so they contribute nothing extra to the solution.}
\vspace{-0.5cm}
\label{fig:68PairChannels}
\end{figure}

The antisymmetrizer $\scA$ for general $N$ is \smash{$\mathcal{S}_{N/2}\otimes \mathcal{S}_{N/2}$}, where $\mathcal{S}$ is the symmetric group. This amounts to \smash{$[(N/2)!]^2$} permutation terms with associated minus signs for each exchange of two identical particles. By applying this operator we find that there is only one H-type channel for \smash{$N=6$} and two H-type channels for \smash{$N=8$}; these are shown in \fref{68PairChannels}. For \smash{$N=6$}, \fref{68PairChannels}(a) shows the H-type channel representing the correlation between a tetramer and an additional pair. Figures \ref{fig:68PairChannels}(b) and \ref{fig:68PairChannels}(c) show, respectively, the \smash{$N=8$} H-type channels with either the correlation between two H-type tetramers, or the correlation between a six-body subcluster and an additional pair. All three include the internal correlations of the smaller subclusters. Below we show that these channels are sufficient for calculating the groundstate of the six- and eight-fermion problem to high accuracy.

Even with only \smash{\smash{$N-1$}} variational parameters in one or two channels it is still not practical with available computational resources to variationally optimize the entire problem when \smash{$N>6$}. However, another advantage to identifying coordinates with correlations is that the basis does not require optimization with respect to the $N$-body problem. Instead, each NIC and IPC is optimized as an independent \smash{$N=2$} subsystem of reduced mass $\mu$. The energy spectrum without interactions and at unitarity is known exactly for \smash{$N=2$} \cite{Busch1998}, and it is simple to optimize these pairs with a gaussian basis and a gaussian potential, \eref{GaussianPotential}. This not only makes the optimization procedure extremely fast, but only a very small number of basis states is needed in each correlation to reproduce the first few energy levels to high accuracy. This allows us to use a very small basis focused on the most important length scales, $r_0$ and $\aho$, and extend the approach to larger $N$. Specifically, the NICs are only of order the trap size so require a smaller basis than the IPCs which also need to access the interparticle potential, i.e.~contain terms with \smash{$\alpha_j\sim r_0$}. 
One drawback to using such a small basis in the \smash{$N=2$} subsystems is that it limits how small we can choose $r_0$ and thus how accurately we can extrapolate \smash{$r_0/\aho \to 0$}. In practice however, the formal requirement that \smash{$r_0/\aho \ll 1$} is satisfied by \smash{$r_0/\aho \lesssim 0.1$} with only a small error compared to the true zero-range limit.

The method is variational even though it does not seek to find convergence of the solution at the $N$-body level. The flexibility and accuracy of this method comes from including all the important correlations in the choice of coordinates. We still use as large a basis size as possible but the terms are distributed more efficiently. The gaussian basis functions are not orthogonal so care must be taken to avoid linear dependence between the differnet \smash{$N=2$} subsystems. However, because the optimization is performed first at the \smash{$N=2$} level, we only need to construct and diagonalize the $N$-body Hamiltonian once, rather than at each step of a full optimization procedure. This leads to a large reduction in computational time especially as the number of permutations grows with $N$.

\subsection{\texorpdfstring{Groundstate energy for \smash{$N=6$}}{N=6} and \texorpdfstring{\smash{$N=8$}}{N=8}}

Similar to the calculation presented in \sref{FourBody} for \smash{$N=4$}, for \smash{$N=6$} we first perform a large calculation to compare against results involving a restricted basis size and choice of channels. Initially, we consider five channels, including several K-type channels, to better incorporate single-particle excitations and the number of basis functions in each correlation is maximized up to the limit of computational resources while maintaining sufficient linear independence. Although still large, the basis size for the ICPs and NICs for \smash{$N=6$} is smaller than for \smash{$N=4$}. Hence, we consider larger values of $r_0/\aho$:  \smash{$0.04\leq r_0/\aho\leq 0.08$}. However, we find that the linear extrapolation as \smash{$r_0/\aho \to 0$} is still valid. In the zero-range limit we obtain a groundstate energy of \smash{$E_\text{G}^{(6)}=6.84(9)\hbar\omega$}, which is lower than the calculation using all two-particle correlations \footnote{Blume and Daily \cite{Blume2011} report a lower figure of \smash{$E_\text{G}^{(6)}=6.842 \hb\omega$} by also extrapolating the basis size to infinity, however this is not strictly in accordance with the variational principle.} and so is a lower upper bound on the true groundstate energy. This is principally because although the basis sizes are comparable we have distributed the basis among only the most important correlations so it accesses a large part of the relevant Hilbert space. Secondly, we repeat the calculation using only the H-type channel and obtain \smash{$E_\text{G}^{(6)}=6.86(4)\hbar\omega$} in the zero-range limit. 
This differs from the larger calculation by $0.22\%$. This demonstrates that the correlations included in the H-type channels are dominant in the groundstate of the system. Details of these results are also summarized in Table \ref{tab:EnergyResults}.


\begin{table}[b!]
\caption{Relative groundstate energies $E_\text{G}^{(N)}$ and details of each calculation. These values are extrapolated from finite $r_0$ calculations to \smash{$r_0/\aho\to 0$}. `Reduced' basis size is in reference to the largest achievable basis for \smash{$N=8$} (see text). The last column is the slope of the linear fit for the extrapolation to \smash{$r_0/\aho\to 0$}.}
\begin{tabular}{l|ccccc}
\hline \hline
$N$ & $E_\text{G}/\hb\omega$ & channels & basis size & $r_0/\aho$ & slope \\ 
\hline  
4 	& 3.509(6)	& H+Ks  	& full		& $[0.01,0.05]$	& 1.833 \\
	& 3.51(3)	& H 		& full 		& $[0.01,0.05]$	& 1.798 \\
	& 3.53(2)	& H 		& red.	 	& $[0.05,0.09]$	& 1.640 \\
\hline
6 	& 6.84(9)	& H+Ks  	& full 		& $[0.04,0.08]$	& 2.717 \\
	& 6.86(4)	& H  		& full	 	& $[0.04,0.08]$	& 2.678 \\
	& 6.91(2)	& H 		& red.	 	& $[0.05,0.09]$	& 2.650 \\
\hline
8 	& 10.63(1)	& H+H		& red.	 	& $[0.05,0.09]$	& 2.841  \\
\hline \hline
\end{tabular}
\label{tab:EnergyResults}
\end{table}

These results demonstrate that the restriction to only H-type channels using a basis of optimized pairs is valid, so we now extend this method to \smash{$N=8$}. There are two linearly independent H-type channels [see \fref{68PairChannels}(b,c)], which must both be included to access all the two-particle IPCs, and there is a much larger number of permutations. The calculations are performed for larger $r_0$: \smash{$0.05\leq r_0/\aho \leq 0.09$}, and the extrapolation to \smash{$r_0/\aho\to 0$} gives \smash{$E_G^{(8)}=10.63(1)\hbar\omega$}. This result is a lower upper bound to the \smash{$N=8$} groundstate energy from Monte Carlo calculations, \smash{$E_G^{(8)}=11.08(3)\hbar\omega$} \cite{Blume2007}, or an effective interaction method, \smash{$E_G^{(8)}=10.679\hbar\omega$} \cite{Mukherjee2013}. 

Although the total basis size for \smash{$N=8$} is maximized, the large number of correlations in the problem means that, in comparison to \smash{$N=4$} and \smash{$N=6$}, we can only use a small number of gaussians in each \smash{$N=2$} subsystem due to limited computational resources. This also means we cannot compare to a calculation that includes K-type channels. The basis is `reduced' at the \smash{$N=2$} level in that it is restricted to the bare minimum needed to reproduce the cusp behavior of the IPCs and the NICs are taken to be of similar size to the trap length, i.e.~\smash{$\alpha_j \sim O(1)$}. The `reduced' basis also requires larger values of $r_0$ in order to maintain similar accuracy in the optimization of the IPCs and NICs. Previous calculations for \smash{$N=4$} and \smash{$N=6$} used a `full' basis size, which essentially allowed arbitrary accuracy in the two-body problems used for optimization. To justify using a `reduced' basis size we repeated the calculations for \smash{$N=4$} and \smash{$N=6$} using the same number of gaussians in each IPC and NIC of the H-type channel as was used in the \smash{$N=8$} calculation. The results and details of all groundstate calculations are summarized in Table \ref{tab:EnergyResults}.
Using the reduced basis size for \smash{$N=4$} and \smash{$N=6$} gives groundstate energies, in the zero-range limit, of \smash{$E_G^{(4)}=3.53(2)\hbar\omega$} and \smash{$E_G^{(6)}=6.91(2)\hbar\omega$}, respectively. The results are within $0.61\%$ and $0.62\%$, respectively, of the `full' basis results. We expect that the error in $E_G^{(8)}$ in the zero-range limit is of similar order.

\section{Structural properties}
\label{sec:Structural}

The coupled pair approach enables the calculation of the structural properties for up to \smash{$N=8$}. In \fref{densities}(a) we plot the normalized single-spin species reduced density \smash{$P_1(r)/\aho^{-3}$} at unitarity, for \smash{$N=4,6,8$} using only the H-type channels with the largest possible basis and smallest possible $r_0$ for each case. 
We see the flattening of the small-$r$ density and emergence of a small peak 
at non-zero $r$ for the six and eight-body, which may indicate formation of 
shell structure in the system. The more prominent peak in Ref.~\cite{Blume2011} may be due to the inclusion of single-particle correlations, but we note that it was performed at a larger value of $r_0$.


In \fref{densities}(b) we plot the (scaled) pair correlation function \smash{$4\pi r^2 P_{12}(r)/\aho^{-1}$} at unitarity for the same parameters and basis sizes as \fref{densities}(a). As $N$ increases the main peak at $r_0/\aho\sim1.5$ is enhanced and becomes the main feature. This is expected as the formation of large weakly-bound dimers due to unitary $s$-wave scattering becomes more likely. The value of $P_{12}(r)$ in the \smash{$r\to0$} limit is related to the number of pairs of fermions of opposite spin separated by a small distance $r$, thereby allowing them to interact \cite{Tan2008a,Werner2012,Daily2009}. In the thermodynamic limit this provides a strong link between the physics of few- and many-body systems but for the values of $N$ considered here the shell structure plays a larger role.

The curves given in \fref{densities} are obtained using the best calculation possible for each $N$ in terms of basis size and the value of $r_0$. In particular, for \smash{$N=8$} the value of $r_0/\aho$ is significantly larger than for \smash{$N=4$} and \smash{$N=6$} although it is still considered small. The near vertical lines in \fref{densities}(b) are due to the small finite range effect of non-zero $r_0$. Decreasing $r_0$ produces a small change in the peak values of $P_{1}(r)$ and $P_{12}(r)$, but this increase is bounded in the \smash{$r_0/\aho \to 0$} limit. Increasing the basis size has the same effect. This effect can be seen by calculating the structural properties of the system for \smash{$N=4$} and \smash{$N=6$} for a range of $r_0$ and for the `reduced' basis as determined by the \smash{$N=8$} system.



\section{Conclusion}
\label{sec:Conclusion}

We have accurately calculated the groundstate energy and structural properties of a two-component few-fermion system at unitarity with an even number of particles up to \smash{$N=8$}. The coupled pair approach minimizes computational effort by retaining only the most important two-body correlations in the unitary limit, allowing the extension to a higher number of particles. For \smash{$N=8$} we present a lower upper bound on the groundstate of \smash{$E_G^{(8)}=10.63(1)\hbar\omega$}. 
The two-particle correlation function $P_{12}(r)$ has two peaks at \smash{$r/\aho \to 0$} and \smash{$r/\aho \sim 1.5$}, representing the number of particles coming within the range of the interparticle interaction and the size of the weakly bound dimers, respectively. As expected for a Fermi system at unitarity, the second peak increases with $N$ due to the dominance of the strong $s$-wave interactions.

\begin{figure}[t]
\includegraphics[width=0.9\columnwidth ]{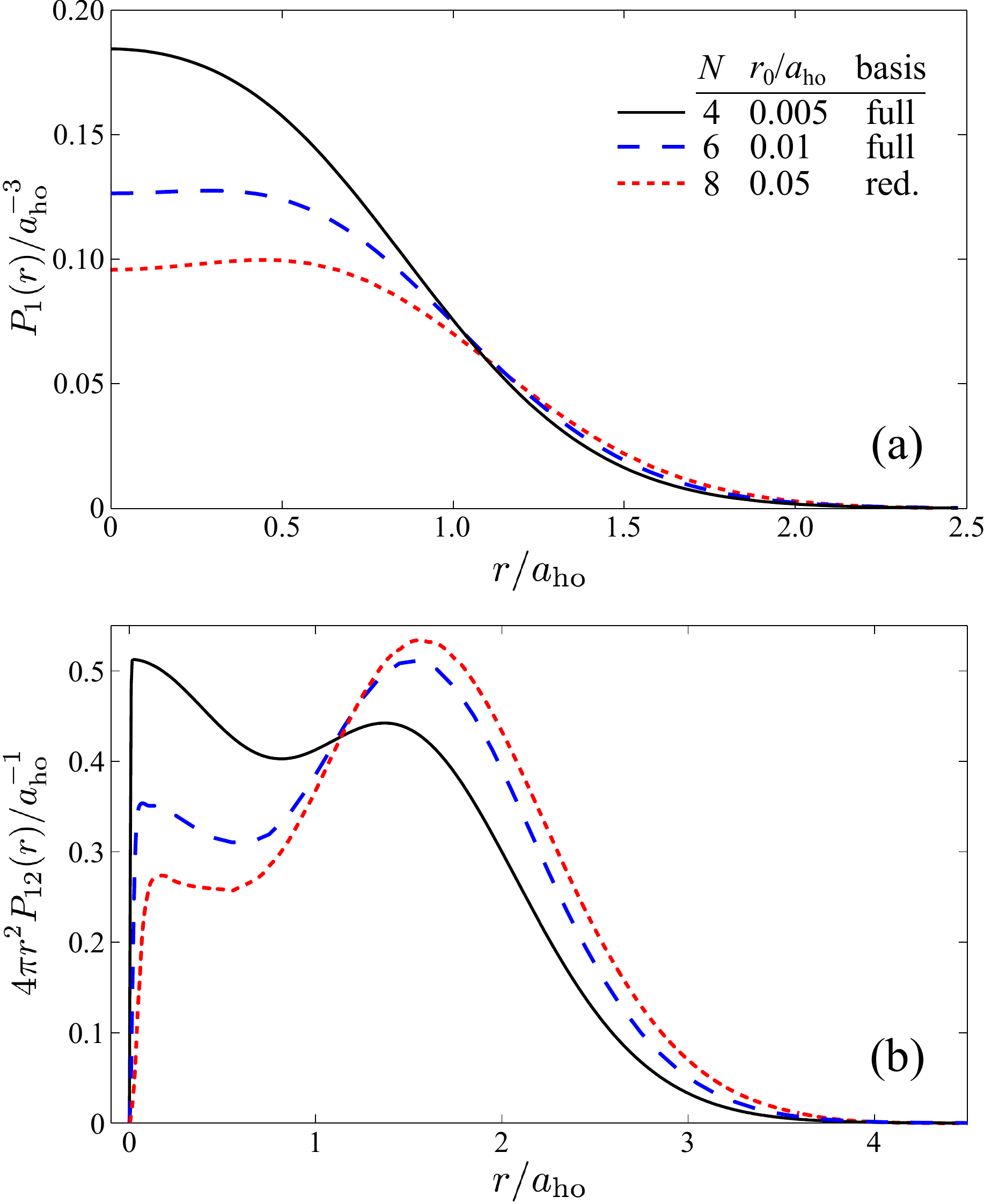}
\caption{(a) Reduced one body density $P_\text{1}(r)/\aho^{-3}$ at unitarity for \smash{$N=4$} (black, solid), \smash{$N=6$} (blue, dashed) and \smash{$N=8$} (red, dotted). In each case calculations were performed using the smallest possible $r_0$ and largest possible basis size. (b) Reduced (and scaled) pair correlation $4\pi r^2 P_\text{12}(r)/\aho^{-1}$ for the same cases.}
\vspace{-0.5cm}
\label{fig:densities}
\end{figure}

All calculations were performed on desktop computers where the principal limiting factor on the computational time for increasing $N$ is the number of permutations required to antisymmetrize the wavefunction. The size of the basis used for each two-body correlation is limited by available memory but even for \smash{$N=8$}, in which at most five gaussians were used for each correlation, the zero-range ground state energy obtained is lower than other techniques. Given these computational limitations, we found that the extrapolation to the zero-range limit was more significant in calculating a lower groundstate energy. Performing the calculation for as small a value of $r_0/\aho$ as possible, while still maintaining high accuracy in the optimization of the IPCs, led to a better linear fit and lower zero-range groundstate energy than increasing the basis size at larger $r_0$.

The ground state of the $N$-fermion problem with equal spin components has zero total angular momentum allowing us to ignore the angular component of the wavefunction in the ansatz of \eref{GaussianMatrixBasis}. The coupled pair approach can be extended to excited states with different symmetry and even other systems with unequal spin populations, odd $N$ or unequal masses. These problems would require the addition of K-type channels and explicit angular functions since single-particle excitations become more important but otherwise the same principles described in this work apply. Advances in the trapping of atomic gases in confined dimensions or with different interparticle interactions \cite{Bloch2008} open up many other avenues in which gaussian expansion and coupled pair methods may apply.

In the unitary limit perturbative many-body techniques fail and new techniques are required. The coupled pair approach is a more efficient method of calculating energetic and structural properties of ultracold Fermi systems with a few atoms. By shifting the computational problem to the most important subsystems of the problem we have highlighted the significance of two-body-correlations as well as pushed the calculation to higher $N$. These results can provide more accurate benchmarks for experiments and calculations in ultracold few-body physics to bridge the gap to the many-body gas.

\begin{acknowledgments}
H.M.Q. acknowledges the support of the ARC Centre of Excellence for Coherent X-ray Science and the ARC Centre of Excellence for Advanced Molecular Imaging.
\end{acknowledgments}

\bibliography{coupled_pair_trapped_fermions} 

\begin{thebibliography}{36}%
\makeatletter
\providecommand \@ifxundefined [1]{%
 \@ifx{#1\undefined}
}%
\providecommand \@ifnum [1]{%
 \ifnum #1\expandafter \@firstoftwo
 \else \expandafter \@secondoftwo
 \fi
}%
\providecommand \@ifx [1]{%
 \ifx #1\expandafter \@firstoftwo
 \else \expandafter \@secondoftwo
 \fi
}%
\providecommand \natexlab [1]{#1}%
\providecommand \enquote  [1]{``#1''}%
\providecommand \bibnamefont  [1]{#1}%
\providecommand \bibfnamefont [1]{#1}%
\providecommand \citenamefont [1]{#1}%
\providecommand \href@noop [0]{\@secondoftwo}%
\providecommand \href [0]{\begingroup \@sanitize@url \@href}%
\providecommand \@href[1]{\@@startlink{#1}\@@href}%
\providecommand \@@href[1]{\endgroup#1\@@endlink}%
\providecommand \@sanitize@url [0]{\catcode `\\12\catcode `\$12\catcode
  `\&12\catcode `\#12\catcode `\^12\catcode `\_12\catcode `\%12\relax}%
\providecommand \@@startlink[1]{}%
\providecommand \@@endlink[0]{}%
\providecommand \url  [0]{\begingroup\@sanitize@url \@url }%
\providecommand \@url [1]{\endgroup\@href {#1}{\urlprefix }}%
\providecommand \urlprefix  [0]{URL }%
\providecommand \Eprint [0]{\href }%
\providecommand \doibase [0]{http://dx.doi.org/}%
\providecommand \selectlanguage [0]{\@gobble}%
\providecommand \bibinfo  [0]{\@secondoftwo}%
\providecommand \bibfield  [0]{\@secondoftwo}%
\providecommand \translation [1]{[#1]}%
\providecommand \BibitemOpen [0]{}%
\providecommand \bibitemStop [0]{}%
\providecommand \bibitemNoStop [0]{.\EOS\space}%
\providecommand \EOS [0]{\spacefactor3000\relax}%
\providecommand \BibitemShut  [1]{\csname bibitem#1\endcsname}%
\let\auto@bib@innerbib\@empty
\bibitem [{\citenamefont {Shin}\ \emph {et~al.}(2008)\citenamefont {Shin},
  \citenamefont {Schunck}, \citenamefont {Schirotzek},\ and\ \citenamefont
  {Ketterle}}]{Shin2008}%
  \BibitemOpen
  \bibfield  {author} {\bibinfo {author} {\bibfnamefont {Y.-i.}\ \bibnamefont
  {Shin}}, \bibinfo {author} {\bibfnamefont {C.~H.}\ \bibnamefont {Schunck}},
  \bibinfo {author} {\bibfnamefont {A.}~\bibnamefont {Schirotzek}}, \ and\
  \bibinfo {author} {\bibfnamefont {W.}~\bibnamefont {Ketterle}},\ }\href
  {http://dx.doi.org/10.1038/nature06473} {\bibfield  {journal} {\bibinfo
  {journal} {Nature}\ }\textbf {\bibinfo {volume} {451}},\ \bibinfo {pages}
  {689} (\bibinfo {year} {2008})}\BibitemShut {NoStop}%
\bibitem [{\citenamefont {Chin}\ \emph {et~al.}(2010)\citenamefont {Chin},
  \citenamefont {Grimm}, \citenamefont {Julienne},\ and\ \citenamefont
  {Tiesinga}}]{Chin2010}%
  \BibitemOpen
  \bibfield  {author} {\bibinfo {author} {\bibfnamefont {C.}~\bibnamefont
  {Chin}}, \bibinfo {author} {\bibfnamefont {R.}~\bibnamefont {Grimm}},
  \bibinfo {author} {\bibfnamefont {P.}~\bibnamefont {Julienne}}, \ and\
  \bibinfo {author} {\bibfnamefont {E.}~\bibnamefont {Tiesinga}},\ }\href
  {http://link.aps.org/doi/10.1103/RevModPhys.82.1225} {\bibfield  {journal}
  {\bibinfo  {journal} {Rev. Mod. Phys.}\ }\textbf {\bibinfo {volume} {82}},\
  \bibinfo {pages} {1225} (\bibinfo {year} {2010})}\BibitemShut {NoStop}%
\bibitem [{\citenamefont {Braaten}\ and\ \citenamefont
  {Hammer}(2006)}]{Braaten2006}%
  \BibitemOpen
  \bibfield  {author} {\bibinfo {author} {\bibfnamefont {E.}~\bibnamefont
  {Braaten}}\ and\ \bibinfo {author} {\bibfnamefont {H.-W.}\ \bibnamefont
  {Hammer}},\ }\href {\doibase http://dx.doi.org/10.1016/j.physrep.2006.03.001}
  {\bibfield  {journal} {\bibinfo  {journal} {Physics Reports}\ }\textbf
  {\bibinfo {volume} {428}},\ \bibinfo {pages} {259 } (\bibinfo {year}
  {2006})}\BibitemShut {NoStop}%
\bibitem [{\citenamefont {Zwierlein}\ \emph {et~al.}(2006)\citenamefont
  {Zwierlein}, \citenamefont {Schunck}, \citenamefont {Schirotzek},\ and\
  \citenamefont {Ketterle}}]{Zwierlein2006}%
  \BibitemOpen
  \bibfield  {author} {\bibinfo {author} {\bibfnamefont {M.~W.}\ \bibnamefont
  {Zwierlein}}, \bibinfo {author} {\bibfnamefont {C.~H.}\ \bibnamefont
  {Schunck}}, \bibinfo {author} {\bibfnamefont {A.}~\bibnamefont {Schirotzek}},
  \ and\ \bibinfo {author} {\bibfnamefont {W.}~\bibnamefont {Ketterle}},\
  }\href {doi:10.1038/nature04936} {\bibfield  {journal} {\bibinfo  {journal}
  {Nature}\ }\textbf {\bibinfo {volume} {442}},\ \bibinfo {pages} {54}
  (\bibinfo {year} {2006})}\BibitemShut {NoStop}%
\bibitem [{\citenamefont {Blume}(2012)}]{Blume2012}%
  \BibitemOpen
  \bibfield  {author} {\bibinfo {author} {\bibfnamefont {D.}~\bibnamefont
  {Blume}},\ }\href {http://stacks.iop.org/0034-4885/75/i=4/a=046401}
  {\bibfield  {journal} {\bibinfo  {journal} {Rep. Prog. Phys.}\ }\textbf
  {\bibinfo {volume} {75}},\ \bibinfo {pages} {046401} (\bibinfo {year}
  {2012})}\BibitemShut {NoStop}%
\bibitem [{\citenamefont {Bloch}\ \emph {et~al.}(2008)\citenamefont {Bloch},
  \citenamefont {Dalibard},\ and\ \citenamefont {Zwerger}}]{Bloch2008}%
  \BibitemOpen
  \bibfield  {author} {\bibinfo {author} {\bibfnamefont {I.}~\bibnamefont
  {Bloch}}, \bibinfo {author} {\bibfnamefont {J.}~\bibnamefont {Dalibard}}, \
  and\ \bibinfo {author} {\bibfnamefont {W.}~\bibnamefont {Zwerger}},\ }\href
  {\doibase 10.1103/RevModPhys.80.885} {\bibfield  {journal} {\bibinfo
  {journal} {Rev. Mod. Phys.}\ }\textbf {\bibinfo {volume} {80}},\ \bibinfo
  {pages} {885} (\bibinfo {year} {2008})}\BibitemShut {NoStop}%
\bibitem [{\citenamefont {Giorgini}\ \emph {et~al.}(2008)\citenamefont
  {Giorgini}, \citenamefont {Pitaevskii},\ and\ \citenamefont
  {Stringari}}]{Giorgini2008}%
  \BibitemOpen
  \bibfield  {author} {\bibinfo {author} {\bibfnamefont {S.}~\bibnamefont
  {Giorgini}}, \bibinfo {author} {\bibfnamefont {L.~P.}\ \bibnamefont
  {Pitaevskii}}, \ and\ \bibinfo {author} {\bibfnamefont {S.}~\bibnamefont
  {Stringari}},\ }\href {\doibase 10.1103/RevModPhys.80.1215} {\bibfield
  {journal} {\bibinfo  {journal} {Rev. Mod. Phys.}\ }\textbf {\bibinfo {volume}
  {80}},\ \bibinfo {pages} {1215} (\bibinfo {year} {2008})}\BibitemShut
  {NoStop}%
\bibitem [{\citenamefont {Gilbreth}\ and\ \citenamefont
  {Alhassid}(2012)}]{Gilbreth2012}%
  \BibitemOpen
  \bibfield  {author} {\bibinfo {author} {\bibfnamefont {C.~N.}\ \bibnamefont
  {Gilbreth}}\ and\ \bibinfo {author} {\bibfnamefont {Y.}~\bibnamefont
  {Alhassid}},\ }\href {\doibase 10.1103/PhysRevA.85.033621} {\bibfield
  {journal} {\bibinfo  {journal} {Phys. Rev. A}\ }\textbf {\bibinfo {volume}
  {85}},\ \bibinfo {pages} {033621} (\bibinfo {year} {2012})}\BibitemShut
  {NoStop}%
\bibitem [{\citenamefont {Blume}\ \emph {et~al.}(2007)\citenamefont {Blume},
  \citenamefont {von Stecher},\ and\ \citenamefont {Greene}}]{Blume2007}%
  \BibitemOpen
  \bibfield  {author} {\bibinfo {author} {\bibfnamefont {D.}~\bibnamefont
  {Blume}}, \bibinfo {author} {\bibfnamefont {J.}~\bibnamefont {von Stecher}},
  \ and\ \bibinfo {author} {\bibfnamefont {C.~H.}\ \bibnamefont {Greene}},\
  }\href {\doibase 10.1103/PhysRevLett.99.233201} {\bibfield  {journal}
  {\bibinfo  {journal} {Phys. Rev. Lett.}\ }\textbf {\bibinfo {volume} {99}},\
  \bibinfo {pages} {233201} (\bibinfo {year} {2007})}\BibitemShut {NoStop}%
\bibitem [{\citenamefont {Mukherjee}\ and\ \citenamefont
  {Alhassid}(2013)}]{Mukherjee2013}%
  \BibitemOpen
  \bibfield  {author} {\bibinfo {author} {\bibfnamefont {A.}~\bibnamefont
  {Mukherjee}}\ and\ \bibinfo {author} {\bibfnamefont {Y.}~\bibnamefont
  {Alhassid}},\ }\href {\doibase 10.1103/PhysRevA.88.053622} {\bibfield
  {journal} {\bibinfo  {journal} {Phys. Rev. A}\ }\textbf {\bibinfo {volume}
  {88}},\ \bibinfo {pages} {053622} (\bibinfo {year} {2013})}\BibitemShut
  {NoStop}%
\bibitem [{\citenamefont {Bulgac}\ and\ \citenamefont {Yu}(2002)}]{Bulgac2002}%
  \BibitemOpen
  \bibfield  {author} {\bibinfo {author} {\bibfnamefont {A.}~\bibnamefont
  {Bulgac}}\ and\ \bibinfo {author} {\bibfnamefont {Y.}~\bibnamefont {Yu}},\
  }\href {\doibase 10.1103/PhysRevLett.88.042504} {\bibfield  {journal}
  {\bibinfo  {journal} {Phys. Rev. Lett.}\ }\textbf {\bibinfo {volume} {88}},\
  \bibinfo {pages} {042504} (\bibinfo {year} {2002})}\BibitemShut {NoStop}%
\bibitem [{\citenamefont {Xianlong}\ \emph {et~al.}(2006)\citenamefont
  {Xianlong}, \citenamefont {Polini}, \citenamefont {Asgari},\ and\
  \citenamefont {Tosi}}]{Xianlong2006}%
  \BibitemOpen
  \bibfield  {author} {\bibinfo {author} {\bibfnamefont {G.}~\bibnamefont
  {Xianlong}}, \bibinfo {author} {\bibfnamefont {M.}~\bibnamefont {Polini}},
  \bibinfo {author} {\bibfnamefont {R.}~\bibnamefont {Asgari}}, \ and\ \bibinfo
  {author} {\bibfnamefont {M.~P.}\ \bibnamefont {Tosi}},\ }\href {\doibase
  10.1103/PhysRevA.73.033609} {\bibfield  {journal} {\bibinfo  {journal} {Phys.
  Rev. A}\ }\textbf {\bibinfo {volume} {73}},\ \bibinfo {pages} {033609}
  (\bibinfo {year} {2006})}\BibitemShut {NoStop}%
\bibitem [{\citenamefont {Papenbrock}(2005)}]{Papenbrock2005}%
  \BibitemOpen
  \bibfield  {author} {\bibinfo {author} {\bibfnamefont {T.}~\bibnamefont
  {Papenbrock}},\ }\href {\doibase 10.1103/PhysRevA.72.041603} {\bibfield
  {journal} {\bibinfo  {journal} {Phys. Rev. A}\ }\textbf {\bibinfo {volume}
  {72}},\ \bibinfo {pages} {041603} (\bibinfo {year} {2005})}\BibitemShut
  {NoStop}%
\bibitem [{\citenamefont {Werner}\ and\ \citenamefont
  {Castin}(2006{\natexlab{a}})}]{Werner2006}%
  \BibitemOpen
  \bibfield  {author} {\bibinfo {author} {\bibfnamefont {F.}~\bibnamefont
  {Werner}}\ and\ \bibinfo {author} {\bibfnamefont {Y.}~\bibnamefont
  {Castin}},\ }\href {\doibase 10.1103/PhysRevLett.97.150401} {\bibfield
  {journal} {\bibinfo  {journal} {Phys. Rev. Lett.}\ }\textbf {\bibinfo
  {volume} {97}},\ \bibinfo {pages} {150401} (\bibinfo {year}
  {2006}{\natexlab{a}})}\BibitemShut {NoStop}%
\bibitem [{\citenamefont {Werner}\ and\ \citenamefont
  {Castin}(2006{\natexlab{b}})}]{Werner2006a}%
  \BibitemOpen
  \bibfield  {author} {\bibinfo {author} {\bibfnamefont {F.}~\bibnamefont
  {Werner}}\ and\ \bibinfo {author} {\bibfnamefont {Y.}~\bibnamefont
  {Castin}},\ }\href {\doibase 10.1103/PhysRevA.74.053604} {\bibfield
  {journal} {\bibinfo  {journal} {Phys. Rev. A}\ }\textbf {\bibinfo {volume}
  {74}},\ \bibinfo {pages} {053604} (\bibinfo {year}
  {2006}{\natexlab{b}})}\BibitemShut {NoStop}%
\bibitem [{\citenamefont {Kestner}\ and\ \citenamefont
  {Duan}(2007)}]{Kestner2007}%
  \BibitemOpen
  \bibfield  {author} {\bibinfo {author} {\bibfnamefont {J.~P.}\ \bibnamefont
  {Kestner}}\ and\ \bibinfo {author} {\bibfnamefont {L.-M.}\ \bibnamefont
  {Duan}},\ }\href {http://link.aps.org/doi/10.1103/PhysRevA.76.033611}
  {\bibfield  {journal} {\bibinfo  {journal} {Phys. Rev. A}\ }\textbf {\bibinfo
  {volume} {76}},\ \bibinfo {pages} {033611} (\bibinfo {year}
  {2007})}\BibitemShut {NoStop}%
\bibitem [{\citenamefont {Daily}\ and\ \citenamefont
  {Blume}(2010)}]{Daily2010}%
  \BibitemOpen
  \bibfield  {author} {\bibinfo {author} {\bibfnamefont {K.~M.}\ \bibnamefont
  {Daily}}\ and\ \bibinfo {author} {\bibfnamefont {D.}~\bibnamefont {Blume}},\
  }\href {http://link.aps.org/doi/10.1103/PhysRevA.81.053615} {\bibfield
  {journal} {\bibinfo  {journal} {Phys. Rev. A}\ }\textbf {\bibinfo {volume}
  {81}},\ \bibinfo {pages} {053615} (\bibinfo {year} {2010})}\BibitemShut
  {NoStop}%
\bibitem [{\citenamefont {Liu}\ \emph {et~al.}(2009)\citenamefont {Liu},
  \citenamefont {Hu},\ and\ \citenamefont {Drummond}}]{Liu2009}%
  \BibitemOpen
  \bibfield  {author} {\bibinfo {author} {\bibfnamefont {X.-J.}\ \bibnamefont
  {Liu}}, \bibinfo {author} {\bibfnamefont {H.}~\bibnamefont {Hu}}, \ and\
  \bibinfo {author} {\bibfnamefont {P.~D.}\ \bibnamefont {Drummond}},\ }\href
  {http://link.aps.org/doi/10.1103/PhysRevLett.102.160401} {\bibfield
  {journal} {\bibinfo  {journal} {Phys. Rev. Lett.}\ }\textbf {\bibinfo
  {volume} {102}},\ \bibinfo {pages} {160401} (\bibinfo {year}
  {2009})}\BibitemShut {NoStop}%
\bibitem [{\citenamefont {Liu}\ \emph {et~al.}(2010)\citenamefont {Liu},
  \citenamefont {Hu},\ and\ \citenamefont {Drummond}}]{Liu2010}%
  \BibitemOpen
  \bibfield  {author} {\bibinfo {author} {\bibfnamefont {X.-J.}\ \bibnamefont
  {Liu}}, \bibinfo {author} {\bibfnamefont {H.}~\bibnamefont {Hu}}, \ and\
  \bibinfo {author} {\bibfnamefont {P.~D.}\ \bibnamefont {Drummond}},\ }\href
  {\doibase 10.1103/PhysRevA.82.023619} {\bibfield  {journal} {\bibinfo
  {journal} {Phys. Rev. A}\ }\textbf {\bibinfo {volume} {82}},\ \bibinfo
  {pages} {023619} (\bibinfo {year} {2010})}\BibitemShut {NoStop}%
\bibitem [{\citenamefont {St\"oferle}\ \emph {et~al.}(2006)\citenamefont
  {St\"oferle}, \citenamefont {Moritz}, \citenamefont {G\"unter}, \citenamefont
  {K\"ohl},\ and\ \citenamefont {Esslinger}}]{Stoeferle2006}%
  \BibitemOpen
  \bibfield  {author} {\bibinfo {author} {\bibfnamefont {T.}~\bibnamefont
  {St\"oferle}}, \bibinfo {author} {\bibfnamefont {H.}~\bibnamefont {Moritz}},
  \bibinfo {author} {\bibfnamefont {K.}~\bibnamefont {G\"unter}}, \bibinfo
  {author} {\bibfnamefont {M.}~\bibnamefont {K\"ohl}}, \ and\ \bibinfo {author}
  {\bibfnamefont {T.}~\bibnamefont {Esslinger}},\ }\href
  {http://link.aps.org/doi/10.1103/PhysRevLett.96.030401} {\bibfield  {journal}
  {\bibinfo  {journal} {Phys. Rev. Lett.}\ }\textbf {\bibinfo {volume} {96}},\
  \bibinfo {pages} {030401} (\bibinfo {year} {2006})}\BibitemShut {NoStop}%
\bibitem [{\citenamefont {Serwane}\ \emph {et~al.}(2011)\citenamefont
  {Serwane}, \citenamefont {Z\"urn}, \citenamefont {Lompe}, \citenamefont
  {Ottenstein}, \citenamefont {Wenz},\ and\ \citenamefont
  {Jochim}}]{Serwane2011}%
  \BibitemOpen
  \bibfield  {author} {\bibinfo {author} {\bibfnamefont {F.}~\bibnamefont
  {Serwane}}, \bibinfo {author} {\bibfnamefont {G.}~\bibnamefont {Z\"urn}},
  \bibinfo {author} {\bibfnamefont {T.}~\bibnamefont {Lompe}}, \bibinfo
  {author} {\bibfnamefont {T.~B.}\ \bibnamefont {Ottenstein}}, \bibinfo
  {author} {\bibfnamefont {A.~N.}\ \bibnamefont {Wenz}}, \ and\ \bibinfo
  {author} {\bibfnamefont {S.}~\bibnamefont {Jochim}},\ }\href {\doibase
  10.1126/science.1201351} {\bibfield  {journal} {\bibinfo  {journal}
  {Science}\ }\textbf {\bibinfo {volume} {332}},\ \bibinfo {pages} {336}
  (\bibinfo {year} {2011})}\BibitemShut {NoStop}%
\bibitem [{\citenamefont {Bakr}\ \emph {et~al.}(2009)\citenamefont {Bakr},
  \citenamefont {Gillen}, \citenamefont {Peng}, \citenamefont {Folling},\ and\
  \citenamefont {Greiner}}]{Bakr2009}%
  \BibitemOpen
  \bibfield  {author} {\bibinfo {author} {\bibfnamefont {W.~S.}\ \bibnamefont
  {Bakr}}, \bibinfo {author} {\bibfnamefont {J.~I.}\ \bibnamefont {Gillen}},
  \bibinfo {author} {\bibfnamefont {A.}~\bibnamefont {Peng}}, \bibinfo {author}
  {\bibfnamefont {S.}~\bibnamefont {Folling}}, \ and\ \bibinfo {author}
  {\bibfnamefont {M.}~\bibnamefont {Greiner}},\ }\href
  {http://dx.doi.org/10.1038/nature08482} {\bibfield  {journal} {\bibinfo
  {journal} {Nature}\ }\textbf {\bibinfo {volume} {462}},\ \bibinfo {pages}
  {74} (\bibinfo {year} {2009})}\BibitemShut {NoStop}%
\bibitem [{\citenamefont {Z\"urn}\ \emph {et~al.}(2013)\citenamefont {Z\"urn},
  \citenamefont {Wenz}, \citenamefont {Murmann}, \citenamefont {Bergschneider},
  \citenamefont {Lompe},\ and\ \citenamefont {Jochim}}]{Zuern2013}%
  \BibitemOpen
  \bibfield  {author} {\bibinfo {author} {\bibfnamefont {G.}~\bibnamefont
  {Z\"urn}}, \bibinfo {author} {\bibfnamefont {A.~N.}\ \bibnamefont {Wenz}},
  \bibinfo {author} {\bibfnamefont {S.}~\bibnamefont {Murmann}}, \bibinfo
  {author} {\bibfnamefont {A.}~\bibnamefont {Bergschneider}}, \bibinfo {author}
  {\bibfnamefont {T.}~\bibnamefont {Lompe}}, \ and\ \bibinfo {author}
  {\bibfnamefont {S.}~\bibnamefont {Jochim}},\ }\href {\doibase
  10.1103/PhysRevLett.111.175302} {\bibfield  {journal} {\bibinfo  {journal}
  {Phys. Rev. Lett.}\ }\textbf {\bibinfo {volume} {111}},\ \bibinfo {pages}
  {175302} (\bibinfo {year} {2013})}\BibitemShut {NoStop}%
\bibitem [{\citenamefont {Mulkerin}\ \emph
  {et~al.}(2012{\natexlab{a}})\citenamefont {Mulkerin}, \citenamefont {Bradly},
  \citenamefont {Quiney},\ and\ \citenamefont {Martin}}]{Mulkerin2012}%
  \BibitemOpen
  \bibfield  {author} {\bibinfo {author} {\bibfnamefont {B.~C.}\ \bibnamefont
  {Mulkerin}}, \bibinfo {author} {\bibfnamefont {C.~J.}\ \bibnamefont
  {Bradly}}, \bibinfo {author} {\bibfnamefont {H.~M.}\ \bibnamefont {Quiney}},
  \ and\ \bibinfo {author} {\bibfnamefont {A.~M.}\ \bibnamefont {Martin}},\
  }\href {\doibase 10.1103/PhysRevA.85.053636} {\bibfield  {journal} {\bibinfo
  {journal} {Phys. Rev. A}\ }\textbf {\bibinfo {volume} {85}},\ \bibinfo
  {pages} {053636} (\bibinfo {year} {2012}{\natexlab{a}})}\BibitemShut
  {NoStop}%
\bibitem [{\citenamefont {Mulkerin}\ \emph
  {et~al.}(2012{\natexlab{b}})\citenamefont {Mulkerin}, \citenamefont {Bradly},
  \citenamefont {Quiney},\ and\ \citenamefont {Martin}}]{Mulkerin2012b}%
  \BibitemOpen
  \bibfield  {author} {\bibinfo {author} {\bibfnamefont {B.~C.}\ \bibnamefont
  {Mulkerin}}, \bibinfo {author} {\bibfnamefont {C.~J.}\ \bibnamefont
  {Bradly}}, \bibinfo {author} {\bibfnamefont {H.~M.}\ \bibnamefont {Quiney}},
  \ and\ \bibinfo {author} {\bibfnamefont {A.~M.}\ \bibnamefont {Martin}},\
  }\href {\doibase 10.1103/PhysRevA.86.053631} {\bibfield  {journal} {\bibinfo
  {journal} {Phys. Rev. A}\ }\textbf {\bibinfo {volume} {86}},\ \bibinfo
  {pages} {053631} (\bibinfo {year} {2012}{\natexlab{b}})}\BibitemShut
  {NoStop}%
\bibitem [{\citenamefont {Busch}\ \emph {et~al.}(1998)\citenamefont {Busch},
  \citenamefont {Englert}, \citenamefont {Rzazewski},\ and\ \citenamefont
  {Wilkens}}]{Busch1998}%
  \BibitemOpen
  \bibfield  {author} {\bibinfo {author} {\bibfnamefont {T.}~\bibnamefont
  {Busch}}, \bibinfo {author} {\bibfnamefont {B.-G.}\ \bibnamefont {Englert}},
  \bibinfo {author} {\bibfnamefont {K.}~\bibnamefont {Rzazewski}}, \ and\
  \bibinfo {author} {\bibfnamefont {M.}~\bibnamefont {Wilkens}},\ }\href
  {\doibase 10.1023/A:1018705520999} {\bibfield  {journal} {\bibinfo  {journal}
  {Found. Phys.}\ }\textbf {\bibinfo {volume} {28}},\ \bibinfo {pages} {549}
  (\bibinfo {year} {1998})}\BibitemShut {NoStop}%
\bibitem [{\citenamefont {Blume}\ and\ \citenamefont
  {Daily}(2011)}]{Blume2011}%
  \BibitemOpen
  \bibfield  {author} {\bibinfo {author} {\bibfnamefont {D.}~\bibnamefont
  {Blume}}\ and\ \bibinfo {author} {\bibfnamefont {K.~M.}\ \bibnamefont
  {Daily}},\ }\href {\doibase doi:10.1016/j.crhy.2010.11.010} {\bibfield
  {journal} {\bibinfo  {journal} {C. R. Phys.}\ }\textbf {\bibinfo {volume}
  {12}},\ \bibinfo {pages} {86} (\bibinfo {year} {2011})}\BibitemShut {NoStop}%
\bibitem [{\citenamefont {Mitroy}\ \emph {et~al.}(2013)\citenamefont {Mitroy},
  \citenamefont {Bubin}, \citenamefont {Horiuchi}, \citenamefont {Suzuki},
  \citenamefont {Adamowicz}, \citenamefont {Cencek}, \citenamefont {Szalewicz},
  \citenamefont {Komasa}, \citenamefont {Blume},\ and\ \citenamefont
  {Varga}}]{Mitroy2013}%
  \BibitemOpen
  \bibfield  {author} {\bibinfo {author} {\bibfnamefont {J.}~\bibnamefont
  {Mitroy}}, \bibinfo {author} {\bibfnamefont {S.}~\bibnamefont {Bubin}},
  \bibinfo {author} {\bibfnamefont {W.}~\bibnamefont {Horiuchi}}, \bibinfo
  {author} {\bibfnamefont {Y.}~\bibnamefont {Suzuki}}, \bibinfo {author}
  {\bibfnamefont {L.}~\bibnamefont {Adamowicz}}, \bibinfo {author}
  {\bibfnamefont {W.}~\bibnamefont {Cencek}}, \bibinfo {author} {\bibfnamefont
  {K.}~\bibnamefont {Szalewicz}}, \bibinfo {author} {\bibfnamefont
  {J.}~\bibnamefont {Komasa}}, \bibinfo {author} {\bibfnamefont
  {D.}~\bibnamefont {Blume}}, \ and\ \bibinfo {author} {\bibfnamefont
  {K.}~\bibnamefont {Varga}},\ }\href {\doibase 10.1103/RevModPhys.85.693}
  {\bibfield  {journal} {\bibinfo  {journal} {Rev. Mod. Phys.}\ }\textbf
  {\bibinfo {volume} {85}},\ \bibinfo {pages} {693} (\bibinfo {year}
  {2013})}\BibitemShut {NoStop}%
\bibitem [{\citenamefont {Suzuki}\ and\ \citenamefont
  {Varga}(1998)}]{Suzuki1998}%
  \BibitemOpen
  \bibfield  {author} {\bibinfo {author} {\bibfnamefont {Y.}~\bibnamefont
  {Suzuki}}\ and\ \bibinfo {author} {\bibfnamefont {K.}~\bibnamefont {Varga}},\
  }\href@noop {} {\emph {\bibinfo {title} {Stochastic Variational Approach to
  Quantum-Mechanical Few-Body Problems}}}\ (\bibinfo  {publisher}
  {Springer-Verlag},\ \bibinfo {year} {1998})\BibitemShut {NoStop}%
\bibitem [{\citenamefont {Hiyama}\ \emph {et~al.}(2003)\citenamefont {Hiyama},
  \citenamefont {Kino},\ and\ \citenamefont {Kamimura}}]{Hiyama2003}%
  \BibitemOpen
  \bibfield  {author} {\bibinfo {author} {\bibfnamefont {E.}~\bibnamefont
  {Hiyama}}, \bibinfo {author} {\bibfnamefont {Y.}~\bibnamefont {Kino}}, \ and\
  \bibinfo {author} {\bibfnamefont {M.}~\bibnamefont {Kamimura}},\ }\href
  {\doibase http://dx.doi.org/10.1016/S0146-6410(03)90015-9} {\bibfield
  {journal} {\bibinfo  {journal} {Progress in Particle and Nuclear Physics}\
  }\textbf {\bibinfo {volume} {51}},\ \bibinfo {pages} {223 } (\bibinfo {year}
  {2003})}\BibitemShut {NoStop}%
\bibitem [{\citenamefont {Rittenhouse}\ \emph {et~al.}(2011)\citenamefont
  {Rittenhouse}, \citenamefont {von Stecher}, \citenamefont {D'Incao},
  \citenamefont {Mehta},\ and\ \citenamefont {Greene}}]{Rittenhouse2011}%
  \BibitemOpen
  \bibfield  {author} {\bibinfo {author} {\bibfnamefont {S.~T.}\ \bibnamefont
  {Rittenhouse}}, \bibinfo {author} {\bibfnamefont {J.}~\bibnamefont {von
  Stecher}}, \bibinfo {author} {\bibfnamefont {J.~P.}\ \bibnamefont {D'Incao}},
  \bibinfo {author} {\bibfnamefont {N.~P.}\ \bibnamefont {Mehta}}, \ and\
  \bibinfo {author} {\bibfnamefont {C.~H.}\ \bibnamefont {Greene}},\ }\href
  {http://stacks.iop.org/0953-4075/44/i=17/a=172001} {\bibfield  {journal}
  {\bibinfo  {journal} {J. Phys. B: At., Mol. Opt. Phys.}\ }\textbf {\bibinfo
  {volume} {44}},\ \bibinfo {pages} {172001} (\bibinfo {year}
  {2011})}\BibitemShut {NoStop}%
\bibitem [{\citenamefont {Petrov}\ \emph {et~al.}(2004)\citenamefont {Petrov},
  \citenamefont {Salomon},\ and\ \citenamefont {Shlyapnikov}}]{Petrov2004}%
  \BibitemOpen
  \bibfield  {author} {\bibinfo {author} {\bibfnamefont {D.~S.}\ \bibnamefont
  {Petrov}}, \bibinfo {author} {\bibfnamefont {C.}~\bibnamefont {Salomon}}, \
  and\ \bibinfo {author} {\bibfnamefont {G.~V.}\ \bibnamefont {Shlyapnikov}},\
  }\href {\doibase 10.1103/PhysRevLett.93.090404} {\bibfield  {journal}
  {\bibinfo  {journal} {Phys. Rev. Lett.}\ }\textbf {\bibinfo {volume} {93}},\
  \bibinfo {pages} {090404} (\bibinfo {year} {2004})}\BibitemShut {NoStop}%
\bibitem [{Note1()}]{Note1}%
  \BibitemOpen
  \bibinfo {note} {Blume and Daily \cite {Blume2011} report a lower figure of
  \protect \smash {$E_\protect \text {G}^{(6)}=6.842 \hbar \omega $} by also
  extrapolating the basis size to infinity, however this is not strictly in
  accordance with the variational principle.}\BibitemShut {Stop}%
\bibitem [{\citenamefont {Tan}(2008)}]{Tan2008a}%
  \BibitemOpen
  \bibfield  {author} {\bibinfo {author} {\bibfnamefont {S.}~\bibnamefont
  {Tan}},\ }\href@noop {} {\bibfield  {journal} {\bibinfo  {journal} {Ann.
  Phys.}\ }\textbf {\bibinfo {volume} {323}},\ \bibinfo {pages} {2971–2986}
  (\bibinfo {year} {2008})}\BibitemShut {NoStop}%
\bibitem [{\citenamefont {Werner}\ and\ \citenamefont
  {Castin}(2012)}]{Werner2012}%
  \BibitemOpen
  \bibfield  {author} {\bibinfo {author} {\bibfnamefont {F.}~\bibnamefont
  {Werner}}\ and\ \bibinfo {author} {\bibfnamefont {Y.}~\bibnamefont
  {Castin}},\ }\href {http://dx.doi.org/ 10.1103/PhysRevA.86.013626} {\bibfield
   {journal} {\bibinfo  {journal} {Phys. Rev. A}\ }\textbf {\bibinfo {volume}
  {86}},\ \bibinfo {pages} {013626} (\bibinfo {year} {2012})}\BibitemShut
  {NoStop}%
\bibitem [{\citenamefont {Daily}\ and\ \citenamefont
  {Blume}(2009)}]{Daily2009}%
  \BibitemOpen
  \bibfield  {author} {\bibinfo {author} {\bibfnamefont {K.~M.}\ \bibnamefont
  {Daily}}\ and\ \bibinfo {author} {\bibfnamefont {D.}~\bibnamefont {Blume}},\
  }\href {http://link.aps.org/doi/10.1103/PhysRevA.80.053626} {\bibfield
  {journal} {\bibinfo  {journal} {Phys. Rev. A}\ }\textbf {\bibinfo {volume}
  {80}},\ \bibinfo {pages} {053626} (\bibinfo {year} {2009})}\BibitemShut
  {NoStop}%
\end{thebibliography}%

\end{document}